\documentclass[aps,twocolumn,prl,showpacs,preprintnumbers,amsmath,amssymb]{revtex4}
\def\etal{{\it et\ al.}}

\newcommand{\lsim} 
 {\ \raise.35ex\hbox{$<$}\kern-0.75em\lower.5ex\hbox{$\sim$}\ }
\newcommand{\gsim}
 {\ \raise.35ex\hbox{$>$}\kern-0.75em\lower.5ex\hbox{$\sim$}\ }

\def\journal #1#2#3#4{#1 {\bf #2}, #3 (#4)}

\def\PRB{Phys.\ Rev.\ B}
\def\PRL{Phys.\ Rev.\ Lett.}

\def\JPSJ{J.\ Phys.\ Soc.\ Jpn.}

\usepackage{graphicx}
\usepackage{dcolumn}
\usepackage{bm}
\usepackage{amsmath}
\usepackage{times}
\usepackage[usenames]{color}
\usepackage{natbib}

\begin{document}
\title{Ring-exchange interaction in doubly degenerate orbital system}
\author{Joji~Nasu, and Sumio~Ishihara}
\affiliation{Department of Physics, Tohoku University, Sendai 980-8578, Japan \\
}
\date{\today}
\begin{abstract}
Ring-exchange (RE) interaction in Mott insulator with doubly degenerate $e_g$ orbitals is studied.  Hamiltonian for the RE interaction is derived by the perturbational calculation. 
A remarkably weak RE interaction destroys the orbital order 
caused by the order-by-fluctuation mechanism in the nearest-neighbor exchange-interaction model. 
A long range order of magnetic octupole moment, i.e. complex orbital wave functions, appears. 
Competition between octupole and quadrupole interactions strongly suppresses ordered moments. 
\end{abstract}


\maketitle



%
%

%




Orbital degree of freedom is one of the central ingredients in Mott insulator and 
correlated metal at vicinity of metal-insulator transition. 
This issue has attracted much attention not only from material viewpoint~\cite{book,JPSJ}, such as colossal-magnetoresisive manganite, superconducting iron punictides, but also from basic theoretical aspects, such as SU(4) spin model~\cite{Li}, and  quantum information~\cite{Kitaev}. 
In spite of recent great effort, fundamental properties in Mott insulator with orbital degree are still under examination. One of the reasons is its highly frustrated character. Since orbital degree implies anisotropy of the electronic wave function, inter-site orbital interaction explicitly depends on a bond direction in a crystal. Thus, interaction energies for all equivalent bonds cannot be minimized simultaneously. 
This intrinsic orbital frustration effect even without geometrical frustration gives rise to a macroscopic number of degeneracy in classical orbital configurations~\cite{Feiner,Khaliulline,Ishihara1,Ishihara2}. It is known that 
this degeneracy is lifted by thermal and quantum fluctuations, and non-trivial orbital ordered state is realized~\cite{Kubo,Nussinov}.  

Multiple-exchange interaction is another candidate to lift the degeneracy. 
The ring-exchange (RE) interaction have been studied for a long time in solid $^{3}$He~\cite{Thouless} 
and strongly correlated electron systems. 
The RE interaction is much weaker than the nearest-neighboring (NN) exchange interaction in usual correlated electron systems. 
However, the RE interaction plays crucial roles to understand Raman and neutron scattering experiments in high-Tc cuprates and spin ladder systems~\cite{Roger,Schmidt,Brehmer}. 
This interaction is also clue to lift the degeneracy and gives rise to a non-trivial ground state in frustrated magnets~\cite{Bernu,Misguich}. Recently, the RE interaction has been studied from view point of quantum criticality between two competing orders~\cite{Senthil,Low}. 
However, the RE interaction in Mott insulator with orbital degeneracy has been left untouched. 

In this Letter, we report the RE interaction effect in orbital degenerate systems. 
In particular, the spin-less $e_g$ orbital model is mainly examined. 
The RE interaction Hamiltonian is derived by the perturbational method. We analyze the Hamiltonian in the classical orbital state at finite temperature 
and the quantum state at zero temperature. 
It is shown that a remarkably weak RE interaction destroys the staggered orbital order 
caused by the order-by-fluctuation mechanism in the NN exchange interaction. 
Magnetic octupole (OP), i.e. a complex wave function for the $e_g$ orbitals, is ordered, in contrast to the NN exchange interaction model. 
Ordered moments are largely suppressed due to strong competition between the electrical quadrupole (QP) and magnetic OP interactions. 
The RE interaction in the spin-orbital coupled system is also discussed.

Let us start from the RE interaction in the spin-less orbital model. 
We consider a Mott insulator with the $e_g$ orbital degree where one electron per site occupies one of the two orbitals, $(u,v)=(d_{3z^2-r^2},d_{x^2-y^2})$, in a cubic lattice. 
The effective model is derived from the spin-less Hubbard model with the $e_g$ orbitals 
defined by 
\begin{eqnarray}
{\cal H}=\sum_{\langle  ij \rangle_a \gamma \gamma'}
\left ( t_a^{\gamma \gamma'} c_{i \gamma}^\dagger c_{j \gamma'} +H.c. \right )
+U\sum_i n_{i u} n_{i v} , 
\label{eq:hubbard}
\end{eqnarray}
where $c_{i \gamma}$ is an annihilation operator of a spin-less fermion with orbital $\gamma(=u, v)$ at site $i$, $n_{i \gamma}=c^\dagger_{i \gamma}c_{i \gamma}$ is the number operator, 
and $\langle ij \rangle_a$ represents the NN $ij$ pair along a direction $a(=x, y, z)$. 
Matrix elements of $t_a^{\gamma \gamma'}$ are given by the Slater-Koster parameters. We define $t \equiv t_z^{uu}$. 

By using the conventional perturbational calculation up to the fourth order of $t$, the effective orbital Hamiltonian is obtained as 
%
$
{\cal H}_{eff}={\cal H}_2+{\cal H}_4 . 
$
The second order term provides the conventional NN exchange interaction given by 
\begin{eqnarray}
{\cal H}_{2}=J\sum_{\langle ij \rangle_a} \tau^{a}_i \tau^{a}_{j} . 
\label{eq:h2}
\end{eqnarray}
The $e_g$ orbital degree of freedom is described by the pseudo-spin (PS) operator with magnitude of 1/2 defined by ${\bf T}_i=(1/2) \sum_{\gamma \gamma'} c^\dagger_{i \gamma} {\bf \sigma}_{\gamma \gamma'} c_{i \gamma'}$ with the Pauli matrices $\bf \sigma$. 
The operator $\tau_{i}^a$ in Eq.~(\ref{eq:h2}) 
is the bond-depend linear combination of $T^x$ and $T^z$ defined by 
$\tau_i^a=\cos(2 \pi n_a/3) T_i^z -\sin(2 \pi n_a/3) T_i^x$
with a number $(n_x, n_y, n_z)=(1,2,3)$. 
The eigen wave function of $\tau^a_i$ with the eigen value of $+1/2 (-1/2)$ 
is $d_{3a^2-r^2}$ $(d_{b^2-c^2})$ where $(a, b, c)$ are the cyclic permutation of $(x,y,z)$. 
The exchange constant is $J=2t^2/U$. 
It is known that, in ${\cal H}_2$, a macroscopic number of degeneracy exists in the classical orbital configurations due to the intrinsic orbital frustration effect~\cite{Feiner,Khaliulline,Ishihara1,Ishihara2}. 
By taking into account thermal and quantum fluctuations, this degeneracy is lifted and the staggered orbital order at $(\pi, \pi, \pi)$ in the Brillouin zone is realized~\cite{Nussinov,Kubo}. 
The ordered pattern is $(d_{3a^2-r^2}, d_{b^2-c^2})$. 
This orbital interaction in Eq.~(\ref{eq:h2}) is also obtained 
by exchange of the Jahn-Teller phonon~\cite{Okamoto}. 

\begin{figure}[t]
\begin{center}
\includegraphics[width=1.0\columnwidth,clip]{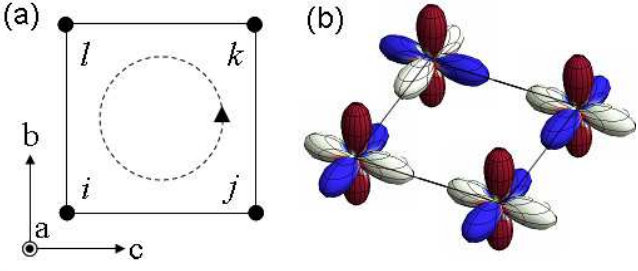}
\caption{(color online) 
Schematic views of (a) the RE interaction, and (b) 
three-up one-down type order of the magnetic OP moment.   
Tones represent phases of the wave functions. 
}
\label{fig:ring}
\end{center}
\end{figure} 
The newly derived fourth-order term of the Hamiltonian is explicitly given by 
\begin{eqnarray}
{\cal H}_4
&=&K_{NN} \sum_{\langle ij \rangle_a}
\left ( \tau_i^a \tau_j^a-{\bar \tau}_i^a {\bar \tau}_j^a-T_i^y T_j^y \right )
\nonumber \\
&+&K_{NNN} \sum_{\langle ij \rangle_{a}}' 
\left ( 
\tau_i^a \tau_j^a-5 {\bar \tau}_i^a {\bar \tau}_j^a+\frac{1}{2}T_i^y T_j^y 
\right ) 
\nonumber \\
&+&K_{3NN} \sum_{\langle ij \rangle_a}''
\tau_i^a \tau_j^a  
+{\cal H}_R , 
\label{eq:h4}
\end{eqnarray}
where we introduce 
${\bar \tau}_i^a=\cos(2 \pi n_a/3) T_i^x + \sin(2 \pi n_a/3) T_i^z$.  
The first, second and third terms of Eq.~(\ref{eq:h4}) are for the interaction 
between the NN $ij$ sites along $a$, 
that between the next NN  sites which is perpendicular to $a$, 
and that between the third NN sites along $a$. 
The exchange constants are 
$K_{NN}=3t^4/(2U^3) $, $K_{NNN}=3t^4/(4U^3)$, and $K_{3NN}=4t^4/U^3$. 
The last term in Eq.~(\ref{eq:h4}) is the RE interaction: 
\begin{eqnarray}
{\cal H}_R=K_{R} \sum_{[ijkl]_a}
\frac{1}{2}
\left ( 
\tau_i^{a +} \tau_j^{a -} \tau_k^{a +} \tau_l^{a -} +H.c. 
\right ),  
\label{eq:ring}
\end{eqnarray}
where 
$\tau^{\pm a}_i=\tau_i^a \pm i (\sqrt{3}/2)T_i^y$ 
and $K_R=40t^4/U^3$. 
A symbol $[i j k l]_a$ represents sites in a plaquette square which is perpendicular to $a$ [see Fig.~\ref{fig:ring}(a)].  
An amplitude of the fourth order terms is represented by a parameter $r_R \equiv \sqrt{K_R/(20J)}$. 
A realistic value of $r_R$ is 0.1-0.2 for LaMnO$_3$ and 0.4-0.5 for LaNiO$_3$, although we examine the RE interaction effect in a wide range of $r_R$. 

Several characteristics in newly obtained ${\cal H}_4$ are listed. 
(i) The RE interaction term, ${\cal H}_R$, is dominant in ${\cal H}_4$ due to its large exchange constant $K_R$. (ii) When we consider ${\cal H}_R$ in a plaquette, the classical stable configuration is that one of the four $\tau_i^{a \pm}$'s is positive (negative) and others are negative (positive). 
This state, termed three-up one-down (3U1D) configuration, 
competes with the staggered orbital configuration favored in ${\cal H}_2$. 
(iii) The $y$-component of the orbital PS $T_i^y$ appears. The eigen states of $T_i^y$ is $(d_u \pm i d_{v})/\sqrt{2}$ which breaks the time-reversal symmetry and represents the magnetic OP with A$_{2g}$ symmetry [see Fig.~\ref{fig:ring}(b)]. 
The $T^y$ operator in ${\cal H}_R$ is caused by the perturabational processes where four plaquette sites are concerned.
This is in contrast to ${\cal H}_2$ where $T_i^x$ and $T_i^z$ implying the electric QP with the $E_g$ symmetry only appears. 
%
%
The OP order was discussed in doped manganites and $f$-electron systems~\cite{Takahashi,Brink,Maezono,Kuramoto}, but not in a Mott insulator. 

\begin{figure}[t]
\begin{center}
\includegraphics[width=0.9\columnwidth,clip]{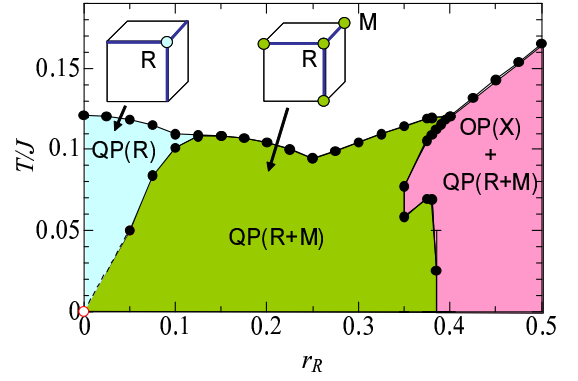}
\caption{(color online) 
Classical phase diagram. Symbols QP(R), QP(R+M) and OP(X)+QP(R+M) represent the staggered QP order, 
the canted QP order and a coexistence of the OP and QP orders, respectively. 
Data at $(T, r_R)=(0,0)$ is obtained by the mean-field method. 
Inset shows degenerate lines at $(T, r_R)=(0, 0)$ and momenta for the QP orders in the Brillouin zone. 
}
\label{fig:classical}
\end{center}
\end{figure} 
First we show the RE interaction effect in the classical orbital state. 
In Fig.~\ref{fig:classical}, the phase diagram obtained by the Monte-Carlo (MC) simulation is presented as a function of $r_R$. 
A three-dimensional $10^3$-site cluster with periodic boundary condition is used. 
We adopt the Wang-Landau algorithm where $5 \times 10^{7}$ MC steps is used for standard measurements. 
The phase diagram is obtained by the correlation functions and the specific heat.
All fourth-order terms are considered, although the calculation in ${\cal H}_2+{\cal H}_R$ qualitatively reproduces the results in the figure. 
As mentioned previously, a macroscopic number of configurations are degenerate at $(r_R, T)=(0,0)$. 
In finite temperatures at $r_R=0$, the staggered QP order
at a momentum $(\pi, \pi, \pi)$ is realized.  
The RE interaction lifts the degeneracy at the point $(r_R, T)=(0, 0)$ and stabilizes 
the canted QP order at $(\pi, \pi, \pi)$ and $(\pi, \pi, 0)$, and other equivalent momenta.
This order is one of the degenerate states at $(r_R, T)=(0,0)$ and competes with the staggered QP order stabilized by the order-by-fluctuation mechanism in ${\cal H}_2$. 
%
It is worth to note that the infinitesimal RE interaction is relevant for the PS configuration, since $(r_R, T)=(0, 0)$ is the degenerate point. 

By further increasing the RE interaction, the OP order appears. 
In this phase, 
the $T^y$ order at $(0, 0, \pi)$ and the non-collinear $T^x-T^z$ order mentioned above  coexist. 
A mechanism of this coexistence phase is as follows. 
As mentioned previously, 
favorable QP ordered patterns in ${\cal H}_J$ and ${\cal H}_R$, i.e. the staggered and 3U1D configurations, respectively, compete with each other. This frustration is released by lifting the PS vector from the $T^x-T^z$ plane toward the $T^y$ axis. 
The cross term $\tau_i^{a} \tau_j^{a} T_k^{y} T_l^{y}$ in ${\cal H}_R$ determines the ordered pattern; in a plaquette, a parallel alignment of $T^y_i T^y_j$ associated with an antiparallel one of $\tau^a_k \tau^a_l$, and vice versa, is stabilized. 

\begin{figure}[t]
\begin{center}
\includegraphics[width=0.7\columnwidth,clip]{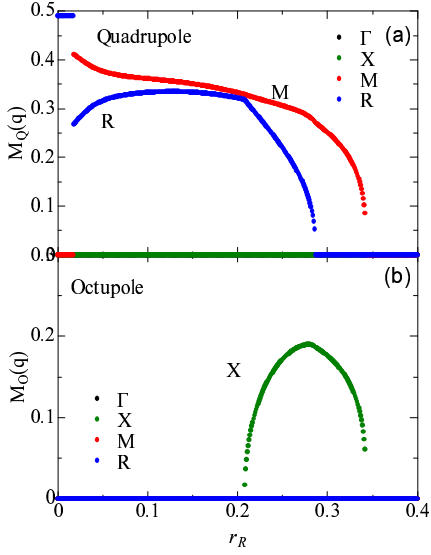}
\caption{(color online) 
(a) Squares of the QP moment, and (b) those in the OP moment obtained by the extended Bethe approximation at representative momenta. 
Data at equivalent momentum in the Brillouin zone are summed. 
}
\label{fig:bethe}
\end{center}
\end{figure} 
Next, we introduce the RE interaction effects in the quantum orbital state at $T=0$. 
The QP and OP ordered moments are obtained by the extended Bethe approximation. 
These are defined by  
$M_Q({\bf q})=[|m^{x}({\bf q})|^2+|m^{z}({\bf q})|^2]^{1/2}$
and 
$M_O({\bf q})=|m^{y}({\bf q})|$, 
respectively, where 
$m^\alpha({\bf q})=(1/N)\sum_i \langle T_i^\alpha \rangle  e^{{\bf q} \cdot {\bf r}_i}$. 
We solve PS states in a $2^3$-site cluster with the mean fields which are obtained self-consistently with the PS states in a cluster. 
The RE interaction in ${\cal H}_R$ is decoupled as 
$\langle \tau_i^{a +} \rangle \langle \tau_j^{a -} \rangle \langle \tau_k^{a +} \rangle \tau_l^{a -}$ at corner sites and as 
$\langle \tau_i^{a +} \tau_j^{a -} \rangle  \tau_k^{a +} \tau_l^{a -}$ in NN bonds in a cluster. We numerically confirmed that this method reproduces the results by the lowest order of the hierarchical mean-field approach proposed in Ref.~\cite{Ortitz}. 
We also apply this method to the spin model consisting of 
the NN exchange and 4-spin RE interactions, and qualitatively reproduce the phase diagram by the exact-diagonalization method in Ref.~\cite{Lauchli} except for the spin nematic phase.   
As shown in Fig.~\ref{fig:bethe}, 
the staggered QP order is stabilized up to a finite value of $r_R$, 
in contrast to the classical phase diagram where the transition occurs at $r_R=0$. 
This is caused by the quantum order-by-fluctuation mechanism where zero-point fluctuation stabilizes the staggered QP order. 
It is worthnoting that the remarkably weak RE interaction ($r_R=0.02$) destroys the staggered QP order and stabilizes the canted QP order favored in the RE interaction. 
With increasing $r_R$, the OP correlation function associated with the QP one 
appears around $r_R=0.22$. 
This phase corresponds to the QP-OP coexistence phase seen in Fig.~\ref{fig:classical}. 
One noticeable point is that both the QP and OP correlation functions vanish at about $r_R=0.35$. This result suggests a suppression of the conventional orders due to quantum fluctuation. 

\begin{figure}[t]
\begin{center}
\includegraphics[width=0.7\columnwidth,clip]{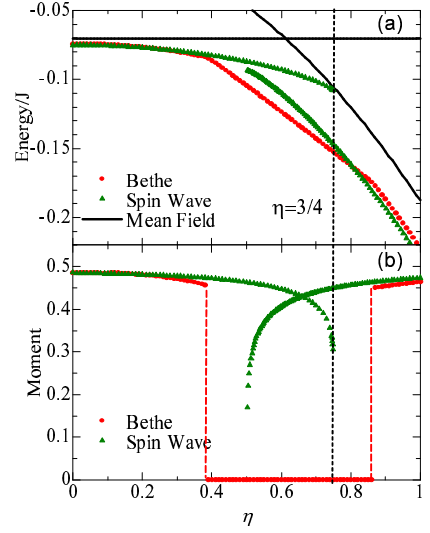}
\caption{(color online) 
(a) Energies and (b) ordered moments in the modified RE interaction Hamiltonian where $\tau_i^{\pm a}$ is replaced by $\tau_i^{\pm a}(\eta)$. 
Circles, triangles and bold lines are for the data obtained by the Bethe method, the spin wave approximation and the mean-field approximation. 
Vertical broken lines indicate $\eta=3/4$.  
}
\label{fig:eta}
\end{center}
\end{figure} 
To clarify the last point furthermore, we study the RE term alone 
and focus on competition between the QP and OP interactions. 
We modify the operator $\tau_i^{a \pm}$ in Eq.~(\ref{eq:ring}) by introducing a parameter $\eta$ as $\tau_i^{\pm a} \rightarrow 
\tau_i^{\pm a}(\eta) \equiv 
\tau_i^a \pm i \sqrt{\eta}T_i^y$. 
It is obvious $\tau_i^{\pm a}(\eta=3/4)=\tau_i^{\pm a}$. 
The energy and the ordered moments, defined by $(\sum_{\alpha q} |m^{\alpha}({\bf q})|^2)^{1/2}$, as a function of $\eta$ are shown in Fig.~\ref{fig:eta}. 
We adopt the extended-Bethe method and the spin-wave approximation where effects of the zero-point vibration is included. 
In the small and large limits of $\eta$, the 3U1D type orders for the QP and OP moments realize, respectively. 
In the results by the spin-wave approximation, 
a point of $\eta=3/4$ is close to the phase boundary between the QP and OP ordered phases and reduction of the ordered moments occur due to competition between the two. 
In the results by the extended Bethe method, this point is inside of the para-orbital phase. 
The wave function in this para-orbital phase is given by the linear combination 
of the 3U1D state. 
We conclude that the suppression of the ordered moments is due to the competition between the OP and QP interactions. 
Higher order calculations are required to judge whether the para-orbital phase survives or not. 

\begin{figure}[t]
\begin{center}
\includegraphics[width=\columnwidth,clip]{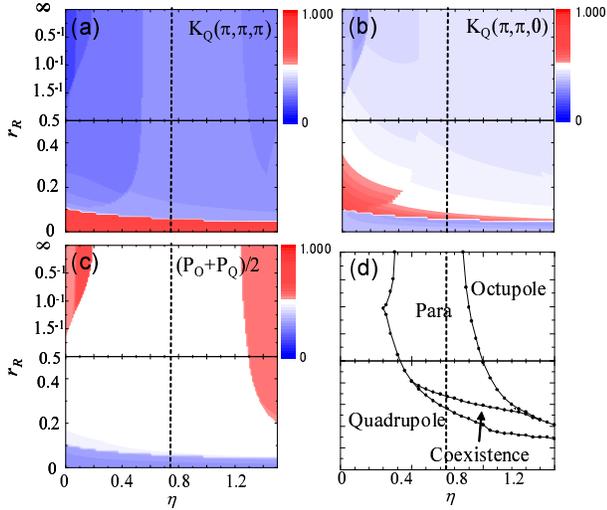}
\caption{(color online) 
A contour map of the two-body correlation function 
$K_Q(\pi,\pi,\pi)$ in (a), that of $K_Q(\pi,\pi,\pi)$ in (b), 
and that of a sum of the QP plaquette correlation functions $P_Q$ and the OP plaquette correlation function $P_O$ in (c). 
Exact diagonalization method in the $2 \times 2 \times 4$ cluster is adopted. 
(d) Phase diagram obtained by the extended Bethe method. 
Vertical broken lines indicate $\eta=3/4$. 
}
\label{fig:ed}
\end{center}
\end{figure} 
Results obtained by the exact diagonalization method helps us to understand whole feature of the phase diagram. 
In Fig.~\ref{fig:ed}, we plot contour maps of the two-body correlation functions $K_Q(\pi,\pi,\pi)$, $K_Q(\pi,\pi,0)$, 
and a sum of the QP plaquette correlation function $P_Q[=(P^x+P^z)/2]$ and 
the OP correlation function $P_O(=P^y)$ in the $r_R-\eta$ plane. 
We introduce 
$K_Q({\bf q})=(4/N^2) \sum_{ij } 
\langle T_i^x  T_j^x+T_i^z T_j^z  \rangle  
e^{{\bf q} \cdot ({\bf r}_i-{\bf r}_j)}$ and 
$P^\alpha=(6N)^{-1}\sum_{[ijkl]}(1-16\langle T_i^\alpha T_j^\alpha T_k^\alpha T_l^\alpha\rangle)$. 
In a region of small $r_R$, the QP orders at $(\pi,\pi,\pi)$ and $(\pi,\pi,0)$ appear.  
On the other side, in large $r_R$, 
large plaquette correlation functions in the regions of small and large $\eta$ 
imply the 3U1D-type QP and OP orders, respectively. 
Among them, there is a wide region where magnitudes of all types of the correlation functions are less than 50$\%$ of their maximum values. 
This region corresponds to the para-orbital phase obtained by the extended Bethe method 
[see Fig.~\ref{fig:ed}(d)]. 
One noticeable point is that region of the para-orbital phase in the Bethe method increases with decreasing $r_R$ from $r_R=\infty$. 
That is, competition between ${\cal H}_2$ and ${\cal H}_R$, as well as the OP and QP interactions in ${\cal H}_R$, contributes to this suppression of the correlation functions. 

Finally, we briefly touch the RE interaction in the spin-orbital coupled system. We derive the RE Hamiltonian from the $e_g$-orbital Hubbard model with spin degree of freedom where the intra-orbital Coulomb interaction ($U_0$), the inter-orbital one ($U'$), the Hund coupling $(J_H)$ and the pair-hopping ($J_P$) are taken into account. 
Since the full spin-orbital effective Hamiltonian up to the fourth order of the transfer integral is much complicated, we briefly introduce the following two cases: (i) the transfer integrals are diagonal, i.e. $t^{\gamma \gamma'}_a=-\delta_{\gamma \gamma'} t$, and (ii) in the intermediate states of the $e_g^2$ configuration in the perturbational processes, the lowest energy state is only considered. 
In the case (i), 
the model at $U_0=U'$ and $J_H=J_P=0$ is known to be the SU(4) point where spin and orbital degrees are equivalent. 
In spite that the transfer integral is diagonal, 
the anisotropic spin order, i.e. the A-type AFM order, is realized between the SU(4) four sublattice phase, favored in the RE interaction, and the ferromagnetic and staggered orbital ordered phase, favored by the Hund coupling. 
In the case (ii), the RE interaction includes a new term which is proportional to ${\bf S}_i \cdot {\bf S}_j \times {\bf S}_k T^y_l$ in a plaquette. 
This term predicts a coexistence of a chiral spin state and a magnetic OP order. 

Summarizing, we have provided a new insight for the orbital physics in correlated electron system beyond the conventional NN exchange interaction, i.e. 
the orbital RE interaction. 
Effects of the RE interaction in the spin-less $e_g$ orbital system are summarized; 
(i) 
a remarkably weak RE interaction destroys the staggered QP order stabilized in ${\cal H}_2$, 
(ii) the magnetic OP order is realized, 
and (iii) large suppression of the ordered moment is caused by the competition of QP and OP 
interactions in ${\cal H}_R$, and that of ${\cal H}_2$ and ${\cal H}_R$.   
As for (iii), we have a theoretical evidence by the extended Bethe approximation, the spin-wave approximation and the exact diagonalization method that the conventional orders are suppressed by introducing the RE interaction. 
However, the issue whether the para-orbital state survives in the higher order calculations or not is left for a future work. 
We believe that the present theory does not only serve a new aspect of the orbital physics and the RE interaction, but also help to understand experiments in Mott insulators near the metal-insulator transition. 

Authors would like to thank M. Matsumoto, Z. Nussinov, and G. Ortitz for their valuable discussions. This work was supported by JSPS KAKENHI, TOKUTEI from MEXT, CREST, and Grand challenges in next-generation integrated nanoscience. JN is supported by the global COE program of MEXT Japan.



\end{document}